\documentclass[
 aip,
 amsmath,amssymb,
 reprint,%
]{revtex4-1}
\usepackage[utf8]{inputenc}
\usepackage[T1]{fontenc}
\usepackage{mathptmx}
\usepackage{etoolbox}
\usepackage{mathptmx}
\usepackage{subfigure}
\usepackage{psfrag}
\usepackage{dcolumn}
\usepackage{bm}
\usepackage{color}
\usepackage{latexsym}
\usepackage{epstopdf}
\usepackage[english]{babel}
\usepackage{times}
\usepackage{latexsym}
\usepackage{natbib}
\usepackage{graphicx}
\usepackage{epsf}
\usepackage{subfigure}
\usepackage{amsmath}                                        
\usepackage{mathtools}
\usepackage{amssymb}
\usepackage{amsfonts}
\usepackage{bm}
\usepackage{appendix}
\usepackage{lipsum, babel}
\usepackage{soul}
\usepackage{hyperref}
\usepackage{braket}
\usepackage{lipsum}
\usepackage{float}
\usepackage{esvect}
\makeatletter

\begin{document}
\title{Four-wave mixing with anti-parity-time symmetry in hot $^{85}$Rb vapor}

\author{Ziqi Niu}
\affiliation{Department of Physics, William \& Mary, Williamsburg, VA 23187, USA}
\author{Yue Jiang}
\affiliation{JILA, National Institute of Standards and Technology and the University of Colorado, Boulder, CO 80309, USA}
\affiliation{Department of Physics, University of Colorado, Boulder, CO 80309, USA}

\author{Jianming Wen}
\affiliation{Department of Physics, Kennesaw State University, Marietta, GA 30060, USA}
\author{Chuanwei Zhang}
\affiliation{Department of Physics, The University of Texas at Dallas, Richardson, TX 75080, USA}
\author{Shengwang Du}
\affiliation{Department of Physics, The University of Texas at Dallas, Richardson, TX 75080, USA}
\author{Irina Novikova}
\affiliation{Department of Physics, William \& Mary, Williamsburg, VA 23187, USA}
\date{\today}

\begin{abstract}
We report an experimental demonstration of anti-parity-time (anti-PT) symmetric optical four-wave mixing in thermal Rubidium vapor, where the propagation of two conjugate optical fields in a double-$\Lambda$ scheme is governed by a non-Hermitian Hamiltonian. We are particularly interested in studying quantum intensity correlations between the two conjugate fields near the exceptional point, taking into account loss and accompanied Langevin noise. Our experimental measurements of classical four-wave mixing gain and the associated two-mode relative-intensity squeezing are in reasonable agreement with the theoretical predictions.
\end{abstract}
\maketitle


While any Hermitian operator has real eigenvalues, being Hermitian is not a necessary condition for this property. Recent demonstrations have shown that any Hamiltonian $\hat{H}$ either symmetric~\cite{bender1998real,bender2007making,el2018non} or anti-symmetric~\cite{ge2013antisymmetric,FanACSPhot2020} under joint parity-time ($\hat{P}\hat{T}$) transformations (with either $[\hat{H},\hat{P}\hat{T}]=0$ or $\{\hat{H},\hat{P}\hat{T}\}=0$, correspondingly), can still yield a real energy spectrum. Both types of systems undergo a phase transition in which the real eigenvalues of the Hamiltonian   become imaginary at a singular point of the parameter space, known as an exceptional point (EP).  Remarkably, even  minimal perturbations of the interaction parameters in the vicinity of the EP may cause dramatic changes in the system observable behavior. This extraordinary sensitivity opens up exciting possibilities for developing sensitive sensors~\cite{ChenNature2017,Wiersig:20,nair2021enhanced,de2021exceptional} and many other applications~\cite{WangAOP23}.
The mathematical equivalence of  Schr\"{o}dinger equation and  paraxial wave propagation equation in materials with complex refractive indices has paved the way for experimental realization of PT and anti-PT symmetric optical and photonic structures by leveraging the spatial variation of their optical properties. Notably, PT-symmetric structures typically employ spatially-interleaved gain and loss channels~\cite{christodoulides2018parity,el2018non,ozdemir2019parity,peng2016anti}, enabling exciting possibilities for practical applications such as EP-enhanced sensing and PT-symmetric lasers~\cite{feng2014single,yu2020experimental,bloch1,laser1,laser2,laser3}. However, unavoidable optical gain and loss pose challenges to many sensing schemes that hold great theoretical promise, since the associated Langevin noises disrupt  PT symmetry in the quantum regime~\cite{zhang2019quantum,naghiloo2019quantum}. Contrarily,
anti-PT symmetric systems offer a promising solution to this issue, as they can potentially be realized without loss or gain by solely manipulating the spatial variation of the real part of refractive indices~\cite{ge2013antisymmetric,li2019anti,bergman2021observation}.

\begin{figure*}
    \centering
    \includegraphics[width=\textwidth]{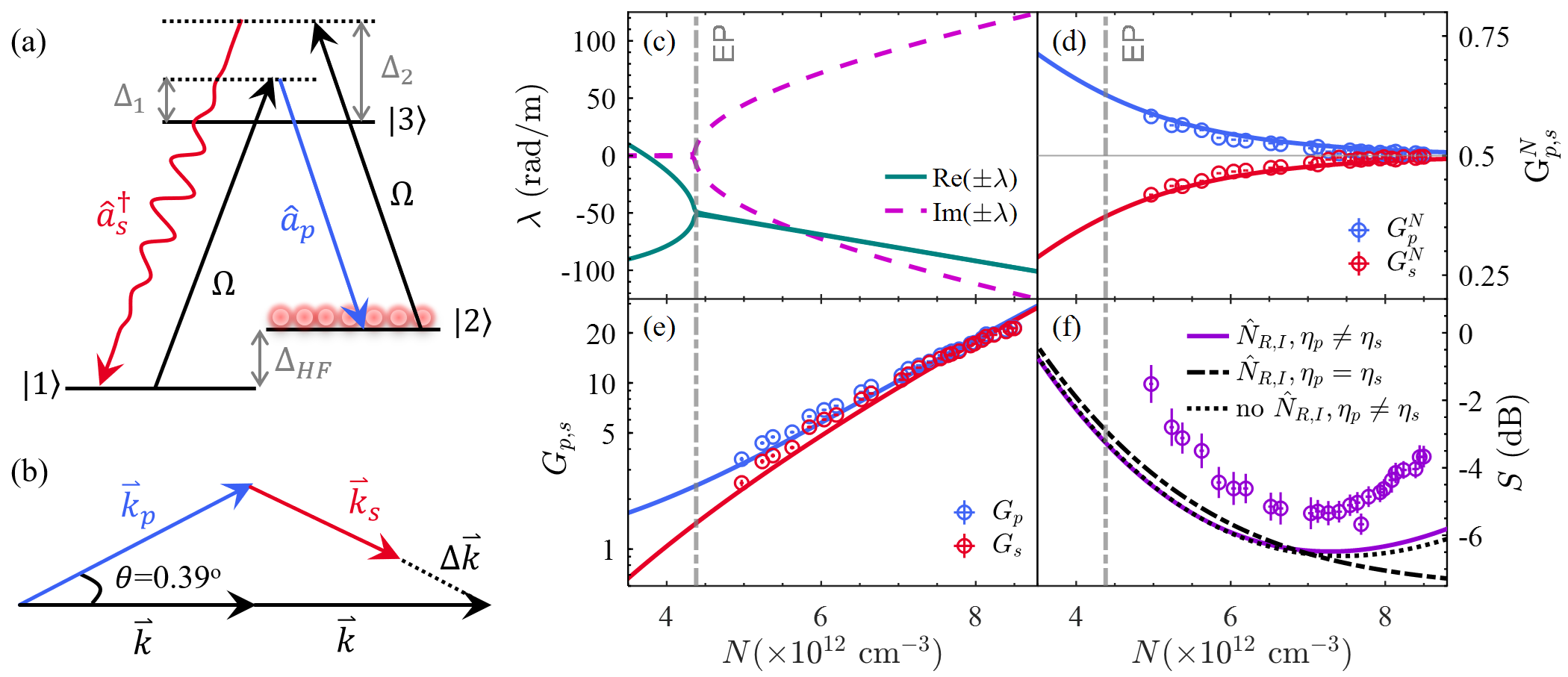}
    \caption{(a) Interaction double-$\Lambda$ scheme used for describing the four-wave mixing process at the D$_1$  transition of $^{85}$Rb. Strong pump laser is detuned by $\Delta_1=0.7$~GHz and $\Delta_2=3.7$~GHz from the $5^2S_{1/2}F=2\rightarrow5^2P_{1/2}F^\prime$ and the $5^2S_{1/2}F=3\rightarrow5^2P_{1/2}F^\prime$ optical transitions, correspondingly. And $\Delta_{HF}=3035.7$ MHz is the ground-state hyperfine splitting. (b) Geometrical arrangement of the optical fields in the FWM process, showing the momentum mismatch $\Delta\vec{k}=2\vec{k}-\vec{k}_p-\vec{k}_s$. (c) Anti-PT Hamiltonian eigenvalues $\pm\lambda$ versus atomic density $N$, calculated using experimental parameters. (d,e) Experimental (markers) and calculated (lines) absolute (d)  and normalized (e)  gain values for the probe and Stokes optical fields versus $N$. 
    (f) Relative intensity squeezing parameter $S$ versus $N$, showing experimental data (markers) and simulated results from the full quantum model with imbalanced (solid line) detector efficiencies  ($\eta_p$ = 78\% and $\eta_s$ = 83\%), from the full quantum model with balanced (dash line) detector efficiencies ($\eta_p = \eta_s$ = 83\%), and from the model included only imbalanced detector loss($\eta_p$ = 78\% and $\eta_s$ = 83\%, 100\% transmission in atomic medium) ~\cite{jasper}(dotted). In all cases the imperfect detector efficiencies are accounted for using a beamsplitter model. 
    For (c)--(e) the dashed vertical lines indicate predicted the EP locations. Experimental parameters: $\theta=0.39^\circ$, pump Rabi frequency $\Omega=2\pi\times0.42$ GHz, cell length $z=1.9$~cm. Temperature range: 100$^\circ$C to $108.7^\circ$C, corresponding to the atomic density range of $N=5\times 10^{12}-9\times 10^{12} \mathrm{cm}^{-3}$. Numerical model used $\Delta k=210$~rad/m extracted from fitting experimental data (see Supplementary Material).}
   
    \label{fig:img1}
\end{figure*}

Recently, a fascinating alternative realization of anti-PT symmetry, without the need for spatially alternating regions with different refractive indices, has been demonstrated in cold Rb atoms~\cite{APTFWM}. In this system, the coupling between two conjugate optical fields is established via resonant four-wave mixing (FWM) with the help of two intense pump laser fields. A nearly lossless propagation and tunable nonlinearity  is achieved  due to strong coupling of light and long-lived ground-state atomic coherence.  By varying the nonlinearity strength, the system exhibited an anti-PT phase transition with the eigenvalues transforming from imaginary to real at the EP. 
Moreover, theory predicts suggest that the quantum fluctuations of the two resulting conjugate optical fields should display distinct behaviors in the vicinity of the EP, which offers a promising platform for precision quantum sensing~\cite{APTSQZ,quant_sense:pooser:2018,PhysRevA.95.063843Lett,ziqi19}. 

Here, we report an experimental realization of a similar anti-PT symmetry in a hot $^{85}$Rb vapor cell using a single strong pump laser field. The system exhibits distinctive anti-PT characteristics in the two conjugate output fields, when the nonlinearity strength is adjusted using atomic density and input pump parameters. We first verified these anti-PT features classically by tracking the FWM gain for both  conjugate fields. Then, we extended our analysis into the quantum realm, analyzing nonclassical correlations in their relative-intensity noise. We consider a more realistic scenario, accounting for residual optical loss and associated Langevin noise, which inevitably reshape the emergence of the anti-PT phase transition and modify squeezing attributes.




We model the four-wave mixing process at the $5^2S_{1/2},F=2,3\rightarrow5^2P_{1/2}$ optical transition of $^{85}$Rb using a three-level double-$\Lambda$ interaction scheme, shown in Fig.~\ref{fig:img1}(a). The pump laser (red) at angular frequency $\omega$ and Rabi frequency $\Omega$ couples atomic transitions $\ket{1}\rightarrow\ket{3}$ and $\ket{2}\rightarrow\ket{3}$ with respective detuning $\Delta_1=0.7$ GHz and $\Delta_2=\Delta_1+\Delta_{HF}=3.7$ GHz, where $\Delta_{HF}=3.035$~GHz is the hyperfine splitting of the $5S_{1/2}$ ground state. The two output conjugate modes, Stokes ($\omega_s,\hat{a}_s^\dagger$) and probe ($\omega_p,\hat{a}_p$), are assumed to only couple to the $\ket{3}\rightarrow\ket{2}$ and $\ket{3}\rightarrow\ket{1}$ transitions, correspondingly, and, under the two-photon resonance condition, to obey the energy conservation $2\omega=\omega_s+\omega_p$. Under the non-depleted pump approximation, the nonlinear FWM interaction between the probe and Stokes field operators, $\hat a_{p}$ and $\hat a_{s}$, is described by the following coupled equations \cite{APTFWM, LangevinPRA2023}
\begin{equation}
i\frac{\partial}{\partial z}\begin{pmatrix}
\hat a_p\\\hat a^\dagger_s
\end{pmatrix}= 
\begin{pmatrix}-\frac{\Delta k}{2}&-\kappa\\
\kappa&\frac{\Delta k}{2}\\
\end{pmatrix}
\begin{pmatrix}
\hat a_p\\\hat a^\dagger_s
\end{pmatrix}.
\label{eqn:propeqnideal1}
\end{equation}
where $\Delta k=2k-(k_{p}+k_{s})\cos\theta$ is the phase mismatch with $k$, $k_p$, and $k_s$ being respectively the wavenumbers of the pump, probe, and Stokes waves and $\theta=0.39^{\circ}$ the fixed misalignment angle (Fig.~\ref{fig:img1}(b)). In simulation we extract the $\Delta k$ by fitting the classical gain data as its value can be effectively modified by the medium.
At the two-photon resonance, $\kappa=gN/2c\Delta_2$ is a real and tunable parametric interaction amplitude with the optical transition coupling strength $g$, 
atomic density $N$, and speed of light in vacuum $c$.

Eq.~\ref{eqn:propeqnideal1}) clearly resembles the the Schr\"{o}dinger-like equation with an effective Hamiltonian:
\begin{equation}\label{eqn:hamiltonianAPT}
\mathbb{H}_{\text{APT}}= \begin{pmatrix}-\frac{\Delta k}{2}&-\kappa\\
\kappa&\frac{\Delta k}{2}\\
\end{pmatrix},
\end{equation}
that anti-commutes with the joint parity-time operator \cite{APTFWM}, 
 $\{\mathbb{H}_{\text{APT}},\hat{P}\hat{T}\}=0$. Since $\mathbb{H}_{\text{APT}}$ (\ref{eqn:hamiltonianAPT}) does not contain any gain or loss, the commutation relations remain intact, removing the need for any  Langevin noise terms in Eq.~(\ref{eqn:propeqnideal1}). 

The Hamiltonian (\ref{eqn:hamiltonianAPT})has two eigenvalues, 
\begin{equation}
\pm\lambda=\pm\frac{\Delta k}{2}\sqrt{1-\beta^2}, 
\end{equation}\label{eqn:eigenvalues}
where $\beta=|2\kappa/\Delta k|$ characterizes standard anti-PT features in parameter space: $\beta=1$ indicates the EP of the regular anti-PT phase transition, marked by both eigenvalue and eigenstate coalescence. For $\beta<1$, $\pm\lambda$ are real, placing the system in the anti-PT phase-broken regime; $\beta>1$ yields imaginary $\pm\lambda$, preserving anti-PT symmetry.
In addition to the aforementioned anti-PT phase transition with $\pm\beta$, we also expect sudden variations of quantum properties of the  conjugate optical fields after interaction with the atomic medium. These variations depend on a transfer matrix connecting output fields at $z=L$ to their corresponding inputs at $z=0$,
\begin{equation}
\begin{pmatrix}
\hat a_p(L)\\ \hat a_s^{\dagger}(L)
\end{pmatrix}= 
e^{-i\mathbb{H}_{\text{APT}}L}
\begin{pmatrix}
\hat a_p(0)\\\hat a_s^{\dagger}(0)
\end{pmatrix}
=\begin{pmatrix}A&C^*\\C&A^*
\end{pmatrix}\begin{pmatrix}
\hat a_p(0)\\\hat a_s^{\dagger}(0)
\end{pmatrix},
\label{eqn:gain1}
\end{equation}
where $A=\cos(\lambda L)+i\sin(\lambda L)/\sqrt{1-\beta^2}$ and $C=-i\beta\sin(\lambda L)/\sqrt{1-\beta^2}$ with $|A|^2-|C|^2=1$. In what follows, we focus on two experimentally-observed parameters: the gain coefficients and relative-intensity fluctuations of the strongly correlated probe and Stokes fields for the different $\beta$ values. 

Let us first examine the classical traits of anti-PT behavior using probe and Stokes field gain values. We define the gain as the ratio of the measured output power to the input power of the seeded input field: $G_j=\left<\hat n_j(L)\right>/\left<\hat n_{\text{seed}}(0)\right>$ ($j=p,s$), where $\hat n_j=\hat a_j^{\dagger}\hat a_j$ is the photon number operator. For weak probe seeding $\langle\hat{n}_p(0)\rangle$, as per Eq.~(\ref{eqn:gain1}), probe and Stokes gains become $G_p=|A|^2$ and $G_s=|C|^2$, respectively. We then can define normalized gains $G_p^N$ and $G^N_s$ as:
\begin{equation}
\begin{aligned}
G_{p}^N&=\frac{G_{p}}{G_p+G_s}=\frac{|A|^2}{|A|^2+|C|^2},\\
G_{s}^N&=\frac{G_{s}}{G_p+G_s}=\frac{|C|^2}{|A|^2+|C|^2}.
\end{aligned}
\end{equation}
For $\beta>1$, both output fields grow exponentially due to the presence of imaginary components in the eigenvalues. Since $|A|^2\approx|C|^2$ for larger $\beta$, the two powers increase at a similar rate, and both $G^N_p$ and $G^N_s$ tend to converge to 0.5. Below the EP ($\beta<1$), coherent power oscillations emerge in both fields. Moreover, as $\beta$ varies, normalized gain for one field increases while that for the other goes down; for $\beta\rightarrow0$, weak FWM strength results in $|A|^2\rightarrow1$ and $|C|^2\rightarrow0$.

Given that the FWM process introduces photons in pairs to both probe and Stokes fields, their intensity fluctuations are expected to be correlated~\cite{lettPRA08,lettSci08,quant_sense:pooser:2018,LettOE2019,Prajapati:19}. Consequently, the quantum aspect of the anti-PT system is revealed in the diminished relative intensity fluctuations between the probe and Stokes fields \cite{APTSQZ}. This reduction is quantified by the squeezing parameter $S$, that  is determined as a relative variance of the probe-Stokes intensity difference at $z=L$ \cite{jasper}, normalized to their shot noise:
\begin{equation}
S = \frac{\textrm{Var}\left(\hat n_p-\hat n_s\right)}{\left<\hat n_p\right>+\left<\hat n_s\right>}=\frac{1}{|A|^2+|C|^2},
\label{eqn:SqzParameter}
\end{equation}
with detailed calculations provided in the Supplementary Material (SM). In this ideal case, it is easy to predict the quantum noise behavior.  When $\beta<1$, $S$ follows sinusoidal oscillations of the classical relative gains, occasionally dropping below the shot noise (when the output powers of the probe and stokes fields become equal), indicating the emergence of moderate quantum squeezing. However, when $\beta>1$, $S$ monotonically decreases, implying growing quantum correlations in relative photon-number fluctuations. A larger $\kappa$ corresponds to better intensity squeezing. Near the EP, $S$ can display rapid variations as $\beta \rightarrow 1$, offering intriguing opportunities for quantum sensing \cite{APTSQZ}. In practice, however, optical loss and imperfect detection efficiency limit the achievable squeezing level, and any further growth in $\kappa$ only leads to deterioration of squeezing and eventually excess noise. Thus, to capture the experimental realities, we develop a model that incorporates the effects of these imperfections, as detailed in the SM.


 To control the value of $\beta$ in our experiment, we choose to vary $\kappa$  by manipulating either the atomic density $N$ or the the Rabi frequency $\Omega$ of the pump laser, keeping $\Delta k$ fixed.  The details of the experimental setup are provided in the Supplementary and in \cite{ziqi19,Prajapati19OL}. In this work, most experimental parameters--such as pump laser frequency and Rabi frequency, and two-photon detuning $\delta = \omega-\omega_p$ have been optimized to maximize the relative-intensity squeezing for different atomic densities. Parameters used for the numerical simulations  are derived from independent experimental characterizations. Since in our model we do not take into account the detailed hyperfine structure of $^{85}$Rb D$_1$ line, the theoretically predicted values for the two optimal two-photon detunings ($\delta\approx-28$ MHz for maximum squeezing and $-17$ MHz for highest gain) differ from the corresponding experimentally measured ones ($1$ MHz and $12$ MHz). Note, however, that in both cases these values are $11$~MHz apart from each other. 

Figures~\ref{fig:img1}(c)-(f) present the variation of the classical and quantum characteristics of the probe and Stokes fields during the anti-PT phase transition as functions of $N$.
Figure~\ref{fig:img1}(c) show the calculated real (green) and imaginary (purple) components of $\pm\lambda$. 
Under the given experimental conditions, the real part of $\lambda_\pm$ above EP does not completely disappears, as expected in the ideal anti-PT scenario. This non-vanishing deviation is caused by additional imaginary contribution $\alpha$ to the diagonal term of the matrix(see Eq.~(\ref{Eq:PropEqMatrix2}) below), introduced to account for residual optical losses for the probe field. Nonetheless, this does not fundamentally disrupt the optical field dynamics, and remains insignificant for applications such as sensing. Figures~\ref{fig:img1}(d) and (e) depict the net gains, $G_p=|A|^2$ and $G_s=|C|^2$, along with the normalized gains, $G^N_{p,s}$, for the probe and Stokes fields, respectively. Both numerical simulations and experimental data exhibit close agreement. Notably, we do not observe any oscillations in the conjugate fields power below EP; instead, the output power of the seeded probe field gradually decrease while the generated Stokes field slowly grows. In principle right after the EP, their normalized gains rapidly converge to 0.5 as the two optical fields tend to equate and grow together, signifying the system's transition into the unbroken domain of the anti-PT phase. Before EP, oscillatory conversion between probe and Stokes is anticipated in the low-atomic-density region for small $\kappa$, stemming from spontaneous symmetry breaking. However, observing these periodic oscillations as well as rapid convergence require longer optical path $L$ (as discussed later) or a significantly larger phase mismatch $\Delta k$. Under this conditions the FWM gain below the EP is very low, posing  experimental challenges.

Figure~\ref{fig:img1}(f) presents the experimentally measured relative intensity squeezing parameter $S$ as well as its numerical simulations for various scenarios. The solid line shows the predictions of the full theoretical model
that assumes the experimentally measured imbalance between probe and stokes detection efficiencies. For completeness, we also plot the model predictions for the case of the identical detector efficiency, shown as a dashed line. The dotted line gives the predictions of a simplified calculations ~\cite{jasper} that neglects the atomic Langevin noise correction terms ($\hat{\mathbb{N}}_{R,I}$). 
As previously mentioned, in an ideal case, quantum correlations between the probe and Stokes intensify with increasing $N$. However, as shown here, in reality the squeezing parameter $S$ reaches its optimal value of $\approx 5$~dB at a certain $N$, above which quantum correlations continuously deteriorate. This shift primarily originates from residual optical loss (particularly for the probe field) that increases quantum noise of each individual optical field and detector losses that hamper fully capturing the generated relative intensity squeezing.  

In the numerical model, a non-negligible $-\alpha$ appears in the relevant diagonal term of $\mathbb{H}_{\text{APT}}$ in Eq.~\eqref{eqn:propeqnideal1}, representing the effective loss rate for the probe field amplitude. The presence of $\alpha$ unavoidably modifies the Heisenberg equations of motion Eq.~\eqref{eqn:propeqnideal1}, requiring adding undesirable Langevin noise and rendering them to 

\begin{equation}
\label{Eq:PropEqMatrix2}
i\partial_z\begin{pmatrix}
\hat{a}_p\\ \hat{a}^{\dagger}_s
\end{pmatrix}=
\begin{pmatrix}-\frac{\Delta k}{2}-i\alpha&-\kappa\\
\kappa&\frac{\Delta k}{2}\\
\end{pmatrix}
\begin{pmatrix}
\hat{a}_p\\ \hat{a}^\dagger_s
\end{pmatrix}+ 
{i\hat{\mathbb{N}}_R}\begin{pmatrix}
\hat{f}_p\\\hat{f}^\dagger_s
\end{pmatrix}+
{i\hat{\mathbb{N}}_I}
\begin{pmatrix}
\hat{f}^\dagger_p\\\hat{f}_s
\end{pmatrix},
\end{equation}
where $\hat{\mathbb{N}}_{R}+i\hat{\mathbb{N}}_{I}=\sqrt{2\begin{pmatrix}\mathrm{Re}(\alpha)&\mathrm{Im}(\kappa)\\
-\mathrm{Im}(\kappa)&0
\end{pmatrix}}$ are the noise matrices, and $\hat{f}_{p,s}$ denote the Langevin noise operators \cite{LangevinPRA2023} (see SM for more details). Assuming the residual absorption insignificant, we can solve Eq.~(\ref{Eq:PropEqMatrix2}) and obtain the gain matrix for the mean amplitudes in the form of Eq.~(\ref{eqn:gain1}). We can then solve the propagation equations for quantum operators and obtain the differential photon-number variance in terms of the gain coefficients $|A|$ and $|C|$:
\begin{equation}
\mathrm{Var}(\hat{n}_{s}-\hat{n}_{p}) = (|A|^2-|C|^2)^2\langle\hat{n}_{p}(0)\rangle+ \langle L_N \rangle,
\label{eqn:VarSqz_wNoise}
\end{equation} 
where $\langle\hat{n}_{p}(0)\rangle$ and $\langle L_N \rangle$, respectively, denote the mean photon number of the seeding probe field and the grouped Langevin noise contributions (see SM for more details).


In the ideal lossless case ($|A|^2 -|C|^2 =1$) without additional noise terms, Eq.~(\ref{eqn:VarSqz_wNoise}) matches Eq.~(\ref{eqn:SqzParameter}) as $\langle\hat{n}_p\rangle + \langle\hat{n}_s\rangle = (|A|^2 +|C|^2)\langle\hat{n}_{p}(0)\rangle$. However, higher $N$ enhances probe-field optical loss, leading to increased excess noise (with super-Poisson statistics) in both conjugate optical fields, and hence prevents further squeezing improvements.  Moreover, small  imbalanced detector losses for the probe and Stokes channels ($\eta_p$ = 78\% and $\eta_s$ = 83\%, respectively) further shifts the the conditions for optimal detectable squeezing further towards lower temperature (atomic density). Eventually, relative intensity noise exceeds the shot-noise level, as depicted by the dotted and solid curves in Fig.~\ref{fig:img1}(f). In the case of perfectly balanced detection better squeezing level can be achieved at higher atomic density. As a side note, one can notice that for low FWM gain slightly higher detection losses for the probe field compensate for unity gain difference between probe and stokes fields, and allows for miniscale improvement in the detected squeezing.
Overall, we observe reasonable agreement between the experimental and theoretically predicted squeezing density dependence. An overall $\approx 2$~dB difference between the measured and calculated noise level is observed for which we attribute   to experimental imperfections such as laser drifts, beam self-focusing, and residual pump field leakage. Nevertheless, the model achieves a reasonably accurate prediction for the overall squeezing trend. This preliminary study is focused within the anti-PT symmetric region, since accurate measurements of quantum noise deviations from the shot noise in the low-gain regime was not possible due to technical noises, such as detector dark noise. 

\begin{figure}[h]\centering
\includegraphics[width=\columnwidth]{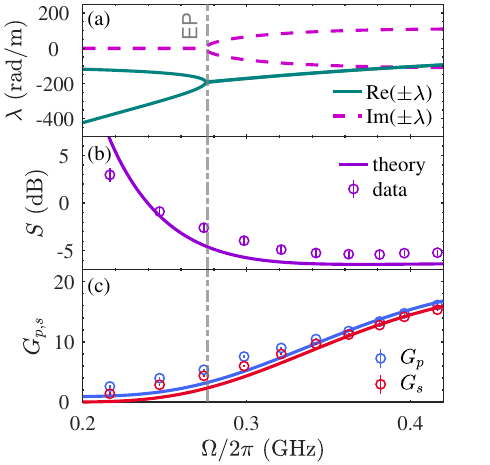}
	\caption{System tuning across anti-PT symmetric and symmetry-breaking regimes via pump laser power modulation. (a) Simulated eigenvalues $\pm\lambda$ of the anti-PT Hamiltonian vs. pump Rabi frequency $\Omega$. Power-dependences  of (b) the relative-intensity squeezing parameter $S$  and (c) normalized probe/Stokes gain $G_{p,s}$; experimental measurements (markers), numerical simulations (lines). Experimental parameters, also used for the numerical model: $\Delta = 0.7$~GHz, $\delta = -28$~MHz, $N = 7.9\times10^{12}\mathrm{cm}^{-3}$ (atomic vapor temperature $\sim$108$^\circ$C), $\Delta k = 210$~rad/m.}%
	\label{fig:fig3sqzrabi}%
\end{figure}


The pump laser power is another experimental parameter that we can use to control the FWM strength. For sufficiently powerful pump field, the FWM gain is independent of the pump laser intensity, but for weaker pump this approach holds potential advantages for much faster tuning across the anti-PT EP, compared to the temperature tuning of the atomic density. Unfortunately, the reduction in pump power generally results in higher optical losses. Figure~\ref{fig:fig3sqzrabi} compares the simulation and experimental results of the pump power dependence. While the experimental normalized gain and measured squeezing align well with the simulations, it is clear that, at lower laser powers, the calculated eigenvalues deviate more substantially from the ideal expectations ($\mathrm{Re}(\pm\lambda)=0$ above the EP, and $\mathrm{Im}(\pm\lambda)=0$ below the EP). 



\begin{figure}[ht]\centering
\includegraphics[width=\columnwidth]{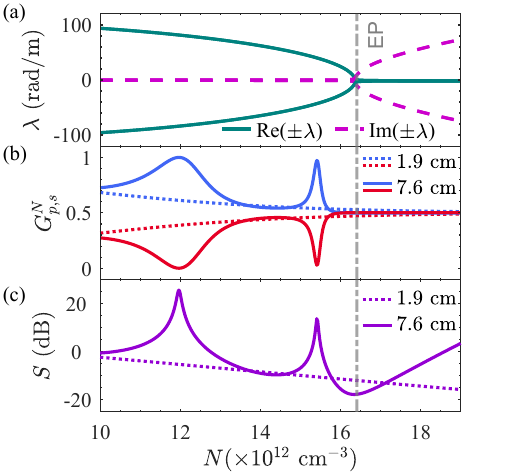} 
	\caption{Optimized FWM parameters for nearly ideal anti-PT realizations, using higher pump power ($\Omega=2\pi\times0.6$ GHz) and larger one-photon detuning ($\Delta_1=4$~GHz), in $z=1.9$ cm and $z=7.6$ cm vapor cells. Additional parameters: $\delta=-3.5$ MHz, $\Delta k=210$ rad/m. (a) Real/imaginary components of the eigenvalues as functions of the atomic density $N$. (b) Normalized probe/Stokes gain vs. $N$. (c) Relative-intensity squeezing parameter $S$ vs. $N$.  Solid/dashed lines show the predicted squeezing with the Langevin noise $\langle L_N\rangle$ at a $z=7.6/1.9$~cm vapor cell, correspondingly.  No detector losses are considered.} 
    \label{fig:anti-PTideal}	
\end{figure}

To demonstrate the capability of our proposed system in simulating near-perfect anti-PT Hamiltonian, we employ the developed numerical model to identify the required experimental conditions, as shown in Fig.~\ref{fig:anti-PTideal}. We find that operating at sufficiently large one-photon detuning $\Delta_{1}\le4$~GHz provides necessary reduction in residual loss. However, to achieve necessary  FWM gain, one will have to operate at higher cell temperature ($\ge 120{}^\circ$C) and greater pump laser power than what was available in the current experiments. 
Under these conditions, the calculated eigenvalues become symmetric and switch from almost entirely real to predominantly imaginary at the EP. For a longer  vapor cell ($z=7.6$ cm), $\pm\lambda z$ attains sufficient magnitude to enable relative oscillation in the normalized gain plot within the anti-PT symmetry breaking region. 
The negligible optical losses make possible to observe corresponding variations in relative-intensity noise below the EP~\cite{APTFWM}, under certain conditions even dipping below the shot-noise level. However, complete elimination of the Langevin noise contributions proves to be challenging. Although in an ideal lossless scenario, squeezing continually improves with $N$, our model predicts that even under more favorable conditions the inescapable optical losses will cause rapid squeezing degradation above certain atomic density as shown in Fig.~\ref{fig:anti-PTideal}(c). 
Operating at larger laser detuning only pushes this optimal squeezing point to higher atomic densities, (compare, e.g., the horizontal scales in Figs.~\ref{fig:img1} and \ref{fig:anti-PTideal}).  
Nevertheless, in the vicinity of the EP, the ability to reproduce rapidly changing quantum squeezing behavior, as identified in Ref.\cite{APTSQZ}, remains feasible.


In conclusion, our preliminary work establishes the practicality of modeling the anti-PT-symmetric Hamiltonian by utilizing two correlated optical fields generated through the near-resonant forward FWM process in hot Rb atoms. We successfully manipulated different experimental parameters to tune the interaction parameters across the anti-PT phase transition and studied both classical and quantum behavior of the probe and Stokes fields. As expected, we observe two-mode relative-intensity squeezing at the anti-PT phase unbroken regime, when both output fields undergo exponential FWM gain. We also analyze the influence of excess noise resulting from residual optical absorption, which imposes constrains on the attainable level of squeezing in distinct domains, and identify reasonable experimental parameters for potential quantum sensor applications. 

This work was supported by the Department of Energy (Grant No. DE-SC0022069); C.Z. acknowledges support from National Science Foundation (PHY-2110212,OMR-2228725). Z.N. and I.N. thank R. Behary and E.E. Mikhailov for their help with the experiment.

\section*{AUTHOR DECLARATIONS}
\textbf{Conflict of Interest}\\
The authors have no conflicts to disclose.

\textbf{Data availability}\\
The data that support the findings of this study are available from the corresponding author upon reasonable request.

\bibliography{bibliography}

\begin{thebibliography}{0}
\expandafter\ifx\csname natexlab\endcsname\relax\def\natexlab#1{#1}\fi
\expandafter\ifx\csname bibnamefont\endcsname\relax
  \def\bibnamefont#1{#1}\fi
\expandafter\ifx\csname bibfnamefont\endcsname\relax
  \def\bibfnamefont#1{#1}\fi
\expandafter\ifx\csname citenamefont\endcsname\relax
  \def\citenamefont#1{#1}\fi
\expandafter\ifx\csname url\endcsname\relax
  \def\url#1{\texttt{#1}}\fi
\expandafter\ifx\csname urlprefix\endcsname\relax\def\urlprefix{URL }\fi
\providecommand{\bibinfo}[2]{#2}
\providecommand{\eprint}[2][]{\url{#2}}

\end{thebibliography}


\begin{thebibliography}{35}%
\makeatletter
\providecommand \@ifxundefined [1]{%
 \@ifx{#1\undefined}
}%
\providecommand \@ifnum [1]{%
 \ifnum #1\expandafter \@firstoftwo
 \else \expandafter \@secondoftwo
 \fi
}%
\providecommand \@ifx [1]{%
 \ifx #1\expandafter \@firstoftwo
 \else \expandafter \@secondoftwo
 \fi
}%
\providecommand \natexlab [1]{#1}%
\providecommand \enquote  [1]{``#1''}%
\providecommand \bibnamefont  [1]{#1}%
\providecommand \bibfnamefont [1]{#1}%
\providecommand \citenamefont [1]{#1}%
\providecommand \href@noop [0]{\@secondoftwo}%
\providecommand \href [0]{\begingroup \@sanitize@url \@href}%
\providecommand \@href[1]{\@@startlink{#1}\@@href}%
\providecommand \@@href[1]{\endgroup#1\@@endlink}%
\providecommand \@sanitize@url [0]{\catcode `\\12\catcode `\$12\catcode
  `\&12\catcode `\#12\catcode `\^12\catcode `\_12\catcode `\%12\relax}%
\providecommand \@@startlink[1]{}%
\providecommand \@@endlink[0]{}%
\providecommand \url  [0]{\begingroup\@sanitize@url \@url }%
\providecommand \@url [1]{\endgroup\@href {#1}{\urlprefix }}%
\providecommand \urlprefix  [0]{URL }%
\providecommand \Eprint [0]{\href }%
\providecommand \doibase [0]{http://dx.doi.org/}%
\providecommand \selectlanguage [0]{\@gobble}%
\providecommand \bibinfo  [0]{\@secondoftwo}%
\providecommand \bibfield  [0]{\@secondoftwo}%
\providecommand \translation [1]{[#1]}%
\providecommand \BibitemOpen [0]{}%
\providecommand \bibitemStop [0]{}%
\providecommand \bibitemNoStop [0]{.\EOS\space}%
\providecommand \EOS [0]{\spacefactor3000\relax}%
\providecommand \BibitemShut  [1]{\csname bibitem#1\endcsname}%
\let\auto@bib@innerbib\@empty
\bibitem [{\citenamefont {Bender}\ and\ \citenamefont
  {Boettcher}(1998)}]{bender1998real}%
  \BibitemOpen
  \bibfield  {author} {\bibinfo {author} {\bibfnamefont {C.~M.}\ \bibnamefont
  {Bender}}\ and\ \bibinfo {author} {\bibfnamefont {S.}~\bibnamefont
  {Boettcher}},\ }\bibfield  {title} {\enquote {\bibinfo {title} {Real spectra
  in non-hermitian hamiltonians having p t symmetry},}\ }\href@noop {}
  {\bibfield  {journal} {\bibinfo  {journal} {Phys. Rev. Lett}\ }\textbf
  {\bibinfo {volume} {80}},\ \bibinfo {pages} {5243} (\bibinfo {year}
  {1998})}\BibitemShut {NoStop}%
\bibitem [{\citenamefont {Bender}(2007)}]{bender2007making}%
  \BibitemOpen
  \bibfield  {author} {\bibinfo {author} {\bibfnamefont {C.~M.}\ \bibnamefont
  {Bender}},\ }\bibfield  {title} {\enquote {\bibinfo {title} {Making sense of
  non-hermitian hamiltonians},}\ }\href@noop {} {\bibfield  {journal} {\bibinfo
   {journal} {Reports on Progress in Physics}\ }\textbf {\bibinfo {volume}
  {70}},\ \bibinfo {pages} {947} (\bibinfo {year} {2007})}\BibitemShut
  {NoStop}%
\bibitem [{\citenamefont {El-Ganainy}\ \emph {et~al.}(2018)\citenamefont
  {El-Ganainy}, \citenamefont {Makris}, \citenamefont {Khajavikhan},
  \citenamefont {Musslimani}, \citenamefont {Rotter},\ and\ \citenamefont
  {Christodoulides}}]{el2018non}%
  \BibitemOpen
  \bibfield  {author} {\bibinfo {author} {\bibfnamefont {R.}~\bibnamefont
  {El-Ganainy}}, \bibinfo {author} {\bibfnamefont {K.~G.}\ \bibnamefont
  {Makris}}, \bibinfo {author} {\bibfnamefont {M.}~\bibnamefont {Khajavikhan}},
  \bibinfo {author} {\bibfnamefont {Z.~H.}\ \bibnamefont {Musslimani}},
  \bibinfo {author} {\bibfnamefont {S.}~\bibnamefont {Rotter}}, \ and\ \bibinfo
  {author} {\bibfnamefont {D.~N.}\ \bibnamefont {Christodoulides}},\ }\bibfield
   {title} {\enquote {\bibinfo {title} {Non-hermitian physics and pt
  symmetry},}\ }\href@noop {} {\bibfield  {journal} {\bibinfo  {journal}
  {Nature Physics}\ }\textbf {\bibinfo {volume} {14}},\ \bibinfo {pages}
  {11--19} (\bibinfo {year} {2018})}\BibitemShut {NoStop}%
\bibitem [{\citenamefont {Ge}\ and\ \citenamefont
  {T{\"u}reci}(2013)}]{ge2013antisymmetric}%
  \BibitemOpen
  \bibfield  {author} {\bibinfo {author} {\bibfnamefont {L.}~\bibnamefont
  {Ge}}\ and\ \bibinfo {author} {\bibfnamefont {H.~E.}\ \bibnamefont
  {T{\"u}reci}},\ }\bibfield  {title} {\enquote {\bibinfo {title}
  {Antisymmetric pt-photonic structures with balanced positive-and
  negative-index materials},}\ }\href@noop {} {\bibfield  {journal} {\bibinfo
  {journal} {Phys. Rev. A}\ }\textbf {\bibinfo {volume} {88}},\ \bibinfo
  {pages} {053810} (\bibinfo {year} {2013})}\BibitemShut {NoStop}%
\bibitem [{\citenamefont {Fan}\ \emph {et~al.}(2020)\citenamefont {Fan},
  \citenamefont {Chen}, \citenamefont {Zhao}, \citenamefont {Wen},\ and\
  \citenamefont {Huang}}]{FanACSPhot2020}%
  \BibitemOpen
  \bibfield  {author} {\bibinfo {author} {\bibfnamefont {H.}~\bibnamefont
  {Fan}}, \bibinfo {author} {\bibfnamefont {J.}~\bibnamefont {Chen}}, \bibinfo
  {author} {\bibfnamefont {Z.}~\bibnamefont {Zhao}}, \bibinfo {author}
  {\bibfnamefont {J.}~\bibnamefont {Wen}}, \ and\ \bibinfo {author}
  {\bibfnamefont {Y.-P.}\ \bibnamefont {Huang}},\ }\bibfield  {title} {\enquote
  {\bibinfo {title} {Antiparity-time symmetry in passive nanophotonics},}\
  }\href@noop {} {\bibfield  {journal} {\bibinfo  {journal} {ACS Photonics}\
  }\textbf {\bibinfo {volume} {7}},\ \bibinfo {pages} {3035--3041} (\bibinfo
  {year} {2020})}\BibitemShut {NoStop}%
\bibitem [{\citenamefont {Chen}\ \emph {et~al.}(2017)\citenamefont {Chen},
  \citenamefont {Şahin Kaya~Özdemir}, \citenamefont {Zhao}, \citenamefont
  {Wiersig},\ and\ \citenamefont {Yang}}]{ChenNature2017}%
  \BibitemOpen
  \bibfield  {author} {\bibinfo {author} {\bibfnamefont {W.}~\bibnamefont
  {Chen}}, \bibinfo {author} {\bibnamefont {Şahin Kaya~Özdemir}}, \bibinfo
  {author} {\bibfnamefont {G.}~\bibnamefont {Zhao}}, \bibinfo {author}
  {\bibfnamefont {J.}~\bibnamefont {Wiersig}}, \ and\ \bibinfo {author}
  {\bibfnamefont {L.}~\bibnamefont {Yang}},\ }\bibfield  {title} {\enquote
  {\bibinfo {title} {Exceptional points enhance sensing in an optical
  microcavity},}\ }\href@noop {} {\bibfield  {journal} {\bibinfo  {journal}
  {Nature}\ }\textbf {\bibinfo {volume} {548}},\ \bibinfo {pages} {192--–196}
  (\bibinfo {year} {2017})}\BibitemShut {NoStop}%
\bibitem [{\citenamefont {Wiersig}(2020)}]{Wiersig:20}%
  \BibitemOpen
  \bibfield  {author} {\bibinfo {author} {\bibfnamefont {J.}~\bibnamefont
  {Wiersig}},\ }\bibfield  {title} {\enquote {\bibinfo {title} {Review of
  exceptional point-based sensors},}\ }\href {\doibase 10.1364/PRJ.396115}
  {\bibfield  {journal} {\bibinfo  {journal} {Photon. Res.}\ }\textbf {\bibinfo
  {volume} {8}},\ \bibinfo {pages} {1457--1467} (\bibinfo {year}
  {2020})}\BibitemShut {NoStop}%
\bibitem [{\citenamefont {Nair}, \citenamefont {Mukhopadhyay},\ and\
  \citenamefont {Agarwal}(2021)}]{nair2021enhanced}%
  \BibitemOpen
  \bibfield  {author} {\bibinfo {author} {\bibfnamefont {J.~M.}\ \bibnamefont
  {Nair}}, \bibinfo {author} {\bibfnamefont {D.}~\bibnamefont {Mukhopadhyay}},
  \ and\ \bibinfo {author} {\bibfnamefont {G.}~\bibnamefont {Agarwal}},\
  }\bibfield  {title} {\enquote {\bibinfo {title} {Enhanced sensing of weak
  anharmonicities through coherences in dissipatively coupled anti-pt symmetric
  systems},}\ }\href@noop {} {\bibfield  {journal} {\bibinfo  {journal}
  {Physical Review Letters}\ }\textbf {\bibinfo {volume} {126}},\ \bibinfo
  {pages} {180401} (\bibinfo {year} {2021})}\BibitemShut {NoStop}%
\bibitem [{\citenamefont {De~Carlo}(2021)}]{de2021exceptional}%
  \BibitemOpen
  \bibfield  {author} {\bibinfo {author} {\bibfnamefont {M.}~\bibnamefont
  {De~Carlo}},\ }\bibfield  {title} {\enquote {\bibinfo {title} {Exceptional
  points of parity-time-and anti-parity-time-symmetric devices for refractive
  index and absorption-based sensing},}\ }\href@noop {} {\bibfield  {journal}
  {\bibinfo  {journal} {Results in Optics}\ }\textbf {\bibinfo {volume} {2}},\
  \bibinfo {pages} {100052} (\bibinfo {year} {2021})}\BibitemShut {NoStop}%
\bibitem [{\citenamefont {Wang}\ \emph {et~al.}(2023)\citenamefont {Wang},
  \citenamefont {Fu}, \citenamefont {Mao}, \citenamefont {Qie}, \citenamefont
  {Stone},\ and\ \citenamefont {Yang}}]{WangAOP23}%
  \BibitemOpen
  \bibfield  {author} {\bibinfo {author} {\bibfnamefont {C.}~\bibnamefont
  {Wang}}, \bibinfo {author} {\bibfnamefont {Z.}~\bibnamefont {Fu}}, \bibinfo
  {author} {\bibfnamefont {W.}~\bibnamefont {Mao}}, \bibinfo {author}
  {\bibfnamefont {J.}~\bibnamefont {Qie}}, \bibinfo {author} {\bibfnamefont
  {A.~D.}\ \bibnamefont {Stone}}, \ and\ \bibinfo {author} {\bibfnamefont
  {L.}~\bibnamefont {Yang}},\ }\bibfield  {title} {\enquote {\bibinfo {title}
  {Non-hermitian optics and photonics: from classical to quantum},}\ }\href
  {\doibase 10.1364/AOP.475477} {\bibfield  {journal} {\bibinfo  {journal}
  {Adv. Opt. Photon.}\ }\textbf {\bibinfo {volume} {15}},\ \bibinfo {pages}
  {442--523} (\bibinfo {year} {2023})}\BibitemShut {NoStop}%
\bibitem [{\citenamefont {Christodoulides}, \citenamefont {Yang}\ \emph
  {et~al.}(2018)\citenamefont {Christodoulides}, \citenamefont {Yang} \emph
  {et~al.}}]{christodoulides2018parity}%
  \BibitemOpen
  \bibfield  {author} {\bibinfo {author} {\bibfnamefont {D.}~\bibnamefont
  {Christodoulides}}, \bibinfo {author} {\bibfnamefont {J.}~\bibnamefont
  {Yang}},  \emph {et~al.},\ }\href@noop {} {\emph {\bibinfo {title}
  {Parity-time symmetry and its applications}}},\ Vol.\ \bibinfo {volume}
  {280}\ (\bibinfo  {publisher} {Springer},\ \bibinfo {year}
  {2018})\BibitemShut {NoStop}%
\bibitem [{\citenamefont {{\"O}zdemir}\ \emph {et~al.}(2019)\citenamefont
  {{\"O}zdemir}, \citenamefont {Rotter}, \citenamefont {Nori},\ and\
  \citenamefont {Yang}}]{ozdemir2019parity}%
  \BibitemOpen
  \bibfield  {author} {\bibinfo {author} {\bibfnamefont {{\c{S}}.~K.}\
  \bibnamefont {{\"O}zdemir}}, \bibinfo {author} {\bibfnamefont
  {S.}~\bibnamefont {Rotter}}, \bibinfo {author} {\bibfnamefont
  {F.}~\bibnamefont {Nori}}, \ and\ \bibinfo {author} {\bibfnamefont
  {L.}~\bibnamefont {Yang}},\ }\bibfield  {title} {\enquote {\bibinfo {title}
  {Parity--time symmetry and exceptional points in photonics},}\ }\href@noop {}
  {\bibfield  {journal} {\bibinfo  {journal} {Nature materials}\ }\textbf
  {\bibinfo {volume} {18}},\ \bibinfo {pages} {783--798} (\bibinfo {year}
  {2019})}\BibitemShut {NoStop}%
\bibitem [{\citenamefont {Peng}\ \emph {et~al.}(2016)\citenamefont {Peng},
  \citenamefont {Cao}, \citenamefont {Shen}, \citenamefont {Qu}, \citenamefont
  {Wen}, \citenamefont {Jiang},\ and\ \citenamefont {Xiao}}]{peng2016anti}%
  \BibitemOpen
  \bibfield  {author} {\bibinfo {author} {\bibfnamefont {P.}~\bibnamefont
  {Peng}}, \bibinfo {author} {\bibfnamefont {W.}~\bibnamefont {Cao}}, \bibinfo
  {author} {\bibfnamefont {C.}~\bibnamefont {Shen}}, \bibinfo {author}
  {\bibfnamefont {W.}~\bibnamefont {Qu}}, \bibinfo {author} {\bibfnamefont
  {J.}~\bibnamefont {Wen}}, \bibinfo {author} {\bibfnamefont {L.}~\bibnamefont
  {Jiang}}, \ and\ \bibinfo {author} {\bibfnamefont {Y.}~\bibnamefont {Xiao}},\
  }\bibfield  {title} {\enquote {\bibinfo {title} {Anti-parity--time symmetry
  with flying atoms},}\ }\href@noop {} {\bibfield  {journal} {\bibinfo
  {journal} {Nature Physics}\ }\textbf {\bibinfo {volume} {12}},\ \bibinfo
  {pages} {1139--1145} (\bibinfo {year} {2016})}\BibitemShut {NoStop}%
\bibitem [{\citenamefont {Feng}\ \emph {et~al.}(2014)\citenamefont {Feng},
  \citenamefont {Wong}, \citenamefont {Ma}, \citenamefont {Wang},\ and\
  \citenamefont {Zhang}}]{feng2014single}%
  \BibitemOpen
  \bibfield  {author} {\bibinfo {author} {\bibfnamefont {L.}~\bibnamefont
  {Feng}}, \bibinfo {author} {\bibfnamefont {Z.~J.}\ \bibnamefont {Wong}},
  \bibinfo {author} {\bibfnamefont {R.-M.}\ \bibnamefont {Ma}}, \bibinfo
  {author} {\bibfnamefont {Y.}~\bibnamefont {Wang}}, \ and\ \bibinfo {author}
  {\bibfnamefont {X.}~\bibnamefont {Zhang}},\ }\bibfield  {title} {\enquote
  {\bibinfo {title} {Single-mode laser by parity-time symmetry breaking},}\
  }\href@noop {} {\bibfield  {journal} {\bibinfo  {journal} {Science}\ }\textbf
  {\bibinfo {volume} {346}},\ \bibinfo {pages} {972--975} (\bibinfo {year}
  {2014})}\BibitemShut {NoStop}%
\bibitem [{\citenamefont {Yu}\ \emph {et~al.}(2020)\citenamefont {Yu},
  \citenamefont {Meng}, \citenamefont {Tang}, \citenamefont {Xu}, \citenamefont
  {Wang}, \citenamefont {Yin}, \citenamefont {Ke}, \citenamefont {Liu},
  \citenamefont {Li}, \citenamefont {Yang} \emph
  {et~al.}}]{yu2020experimental}%
  \BibitemOpen
  \bibfield  {author} {\bibinfo {author} {\bibfnamefont {S.}~\bibnamefont
  {Yu}}, \bibinfo {author} {\bibfnamefont {Y.}~\bibnamefont {Meng}}, \bibinfo
  {author} {\bibfnamefont {J.-S.}\ \bibnamefont {Tang}}, \bibinfo {author}
  {\bibfnamefont {X.-Y.}\ \bibnamefont {Xu}}, \bibinfo {author} {\bibfnamefont
  {Y.-T.}\ \bibnamefont {Wang}}, \bibinfo {author} {\bibfnamefont
  {P.}~\bibnamefont {Yin}}, \bibinfo {author} {\bibfnamefont {Z.-J.}\
  \bibnamefont {Ke}}, \bibinfo {author} {\bibfnamefont {W.}~\bibnamefont
  {Liu}}, \bibinfo {author} {\bibfnamefont {Z.-P.}\ \bibnamefont {Li}},
  \bibinfo {author} {\bibfnamefont {Y.-Z.}\ \bibnamefont {Yang}},  \emph
  {et~al.},\ }\bibfield  {title} {\enquote {\bibinfo {title} {Experimental
  investigation of quantum p t-enhanced sensor},}\ }\href@noop {} {\bibfield
  {journal} {\bibinfo  {journal} {Phys. Rev. Lett}\ }\textbf {\bibinfo {volume}
  {125}},\ \bibinfo {pages} {240506} (\bibinfo {year} {2020})}\BibitemShut
  {NoStop}%
\bibitem [{\citenamefont {Longhi}(2009)}]{bloch1}%
  \BibitemOpen
  \bibfield  {author} {\bibinfo {author} {\bibfnamefont {S.}~\bibnamefont
  {Longhi}},\ }\bibfield  {title} {\enquote {\bibinfo {title} {Bloch
  oscillations in complex crystals with p t symmetry},}\ }\href@noop {}
  {\bibfield  {journal} {\bibinfo  {journal} {Phys. Rev. Lett}\ }\textbf
  {\bibinfo {volume} {103}},\ \bibinfo {pages} {123601} (\bibinfo {year}
  {2009})}\BibitemShut {NoStop}%
\bibitem [{\citenamefont {Ge}\ \emph {et~al.}(2011)\citenamefont {Ge},
  \citenamefont {Chong}, \citenamefont {Rotter}, \citenamefont {T{\"u}reci},\
  and\ \citenamefont {Stone}}]{laser1}%
  \BibitemOpen
  \bibfield  {author} {\bibinfo {author} {\bibfnamefont {L.}~\bibnamefont
  {Ge}}, \bibinfo {author} {\bibfnamefont {Y.}~\bibnamefont {Chong}}, \bibinfo
  {author} {\bibfnamefont {S.}~\bibnamefont {Rotter}}, \bibinfo {author}
  {\bibfnamefont {H.~E.}\ \bibnamefont {T{\"u}reci}}, \ and\ \bibinfo {author}
  {\bibfnamefont {A.}~\bibnamefont {Stone}},\ }\bibfield  {title} {\enquote
  {\bibinfo {title} {Unconventional modes in lasers with spatially varying gain
  and loss},}\ }\href@noop {} {\bibfield  {journal} {\bibinfo  {journal} {Phys.
  Rev. A}\ }\textbf {\bibinfo {volume} {84}},\ \bibinfo {pages} {023820}
  (\bibinfo {year} {2011})}\BibitemShut {NoStop}%
\bibitem [{\citenamefont {Teimourpour}\ \emph {et~al.}(2016)\citenamefont
  {Teimourpour}, \citenamefont {Ge}, \citenamefont {Christodoulides},\ and\
  \citenamefont {El-Ganainy}}]{laser2}%
  \BibitemOpen
  \bibfield  {author} {\bibinfo {author} {\bibfnamefont {M.~H.}\ \bibnamefont
  {Teimourpour}}, \bibinfo {author} {\bibfnamefont {L.}~\bibnamefont {Ge}},
  \bibinfo {author} {\bibfnamefont {D.~N.}\ \bibnamefont {Christodoulides}}, \
  and\ \bibinfo {author} {\bibfnamefont {R.}~\bibnamefont {El-Ganainy}},\
  }\bibfield  {title} {\enquote {\bibinfo {title} {Non-hermitian engineering of
  single mode two dimensional laser arrays},}\ }\href@noop {} {\bibfield
  {journal} {\bibinfo  {journal} {Scientific reports}\ }\textbf {\bibinfo
  {volume} {6}},\ \bibinfo {pages} {1--9} (\bibinfo {year} {2016})}\BibitemShut
  {NoStop}%
\bibitem [{\citenamefont {Hokmabadi}\ \emph {et~al.}(2019)\citenamefont
  {Hokmabadi}, \citenamefont {Nye}, \citenamefont {El-Ganainy}, \citenamefont
  {Christodoulides},\ and\ \citenamefont {Khajavikhan}}]{laser3}%
  \BibitemOpen
  \bibfield  {author} {\bibinfo {author} {\bibfnamefont {M.~P.}\ \bibnamefont
  {Hokmabadi}}, \bibinfo {author} {\bibfnamefont {N.~S.}\ \bibnamefont {Nye}},
  \bibinfo {author} {\bibfnamefont {R.}~\bibnamefont {El-Ganainy}}, \bibinfo
  {author} {\bibfnamefont {D.~N.}\ \bibnamefont {Christodoulides}}, \ and\
  \bibinfo {author} {\bibfnamefont {M.}~\bibnamefont {Khajavikhan}},\
  }\bibfield  {title} {\enquote {\bibinfo {title} {Supersymmetric laser
  arrays},}\ }\href@noop {} {\bibfield  {journal} {\bibinfo  {journal}
  {Science}\ }\textbf {\bibinfo {volume} {363}},\ \bibinfo {pages} {623--626}
  (\bibinfo {year} {2019})}\BibitemShut {NoStop}%
\bibitem [{\citenamefont {Zhang}\ \emph {et~al.}(2019)\citenamefont {Zhang},
  \citenamefont {Sweeney}, \citenamefont {Hsu}, \citenamefont {Yang},
  \citenamefont {Stone},\ and\ \citenamefont {Jiang}}]{zhang2019quantum}%
  \BibitemOpen
  \bibfield  {author} {\bibinfo {author} {\bibfnamefont {M.}~\bibnamefont
  {Zhang}}, \bibinfo {author} {\bibfnamefont {W.}~\bibnamefont {Sweeney}},
  \bibinfo {author} {\bibfnamefont {C.~W.}\ \bibnamefont {Hsu}}, \bibinfo
  {author} {\bibfnamefont {L.}~\bibnamefont {Yang}}, \bibinfo {author}
  {\bibfnamefont {A.}~\bibnamefont {Stone}}, \ and\ \bibinfo {author}
  {\bibfnamefont {L.}~\bibnamefont {Jiang}},\ }\bibfield  {title} {\enquote
  {\bibinfo {title} {Quantum noise theory of exceptional point amplifying
  sensors},}\ }\href@noop {} {\bibfield  {journal} {\bibinfo  {journal} {Phys.
  Rev. Lett}\ }\textbf {\bibinfo {volume} {123}},\ \bibinfo {pages} {180501}
  (\bibinfo {year} {2019})}\BibitemShut {NoStop}%
\bibitem [{\citenamefont {Naghiloo}\ \emph {et~al.}(2019)\citenamefont
  {Naghiloo}, \citenamefont {Abbasi}, \citenamefont {Joglekar},\ and\
  \citenamefont {Murch}}]{naghiloo2019quantum}%
  \BibitemOpen
  \bibfield  {author} {\bibinfo {author} {\bibfnamefont {M.}~\bibnamefont
  {Naghiloo}}, \bibinfo {author} {\bibfnamefont {M.}~\bibnamefont {Abbasi}},
  \bibinfo {author} {\bibfnamefont {Y.~N.}\ \bibnamefont {Joglekar}}, \ and\
  \bibinfo {author} {\bibfnamefont {K.}~\bibnamefont {Murch}},\ }\bibfield
  {title} {\enquote {\bibinfo {title} {Quantum state tomography across the
  exceptional point in a single dissipative qubit},}\ }\href@noop {} {\bibfield
   {journal} {\bibinfo  {journal} {Nature Physics}\ }\textbf {\bibinfo {volume}
  {15}},\ \bibinfo {pages} {1232--1236} (\bibinfo {year} {2019})}\BibitemShut
  {NoStop}%
\bibitem [{\citenamefont {Li}\ \emph {et~al.}(2019)\citenamefont {Li},
  \citenamefont {Peng}, \citenamefont {Han}, \citenamefont {Miri},
  \citenamefont {Li}, \citenamefont {Xiao}, \citenamefont {Zhu}, \citenamefont
  {Zhao}, \citenamefont {Al{\`u}}, \citenamefont {Fan} \emph
  {et~al.}}]{li2019anti}%
  \BibitemOpen
  \bibfield  {author} {\bibinfo {author} {\bibfnamefont {Y.}~\bibnamefont
  {Li}}, \bibinfo {author} {\bibfnamefont {Y.-G.}\ \bibnamefont {Peng}},
  \bibinfo {author} {\bibfnamefont {L.}~\bibnamefont {Han}}, \bibinfo {author}
  {\bibfnamefont {M.-A.}\ \bibnamefont {Miri}}, \bibinfo {author}
  {\bibfnamefont {W.}~\bibnamefont {Li}}, \bibinfo {author} {\bibfnamefont
  {M.}~\bibnamefont {Xiao}}, \bibinfo {author} {\bibfnamefont {X.-F.}\
  \bibnamefont {Zhu}}, \bibinfo {author} {\bibfnamefont {J.}~\bibnamefont
  {Zhao}}, \bibinfo {author} {\bibfnamefont {A.}~\bibnamefont {Al{\`u}}},
  \bibinfo {author} {\bibfnamefont {S.}~\bibnamefont {Fan}},  \emph {et~al.},\
  }\bibfield  {title} {\enquote {\bibinfo {title} {Anti--parity-time symmetry
  in diffusive systems},}\ }\href@noop {} {\bibfield  {journal} {\bibinfo
  {journal} {Science}\ }\textbf {\bibinfo {volume} {364}},\ \bibinfo {pages}
  {170--173} (\bibinfo {year} {2019})}\BibitemShut {NoStop}%
\bibitem [{\citenamefont {Bergman}\ \emph {et~al.}(2021)\citenamefont
  {Bergman}, \citenamefont {Duggan}, \citenamefont {Sharma}, \citenamefont
  {Tur}, \citenamefont {Zadok},\ and\ \citenamefont
  {Al{\`u}}}]{bergman2021observation}%
  \BibitemOpen
  \bibfield  {author} {\bibinfo {author} {\bibfnamefont {A.}~\bibnamefont
  {Bergman}}, \bibinfo {author} {\bibfnamefont {R.}~\bibnamefont {Duggan}},
  \bibinfo {author} {\bibfnamefont {K.}~\bibnamefont {Sharma}}, \bibinfo
  {author} {\bibfnamefont {M.}~\bibnamefont {Tur}}, \bibinfo {author}
  {\bibfnamefont {A.}~\bibnamefont {Zadok}}, \ and\ \bibinfo {author}
  {\bibfnamefont {A.}~\bibnamefont {Al{\`u}}},\ }\bibfield  {title} {\enquote
  {\bibinfo {title} {Observation of anti-parity-time-symmetry, phase
  transitions and exceptional points in an optical fibre},}\ }\href@noop {}
  {\bibfield  {journal} {\bibinfo  {journal} {Nature communications}\ }\textbf
  {\bibinfo {volume} {12}},\ \bibinfo {pages} {1--9} (\bibinfo {year}
  {2021})}\BibitemShut {NoStop}%
\bibitem [{\citenamefont {Jasperse}(2010)}]{jasper}%
  \BibitemOpen
  \bibfield  {author} {\bibinfo {author} {\bibfnamefont {M.}~\bibnamefont
  {Jasperse}},\ }\bibfield  {title} {\enquote {\bibinfo {title} {Relative
  intensity squeezing: by four-wave mixing in rubidium},}\ }\href@noop {} {\
  (\bibinfo {year} {2010})}\BibitemShut {NoStop}%
\bibitem [{\citenamefont {Jiang}\ \emph {et~al.}(2019)\citenamefont {Jiang},
  \citenamefont {Mei}, \citenamefont {Zuo}, \citenamefont {Zhai}, \citenamefont
  {Li}, \citenamefont {Wen},\ and\ \citenamefont {Du}}]{APTFWM}%
  \BibitemOpen
  \bibfield  {author} {\bibinfo {author} {\bibfnamefont {Y.}~\bibnamefont
  {Jiang}}, \bibinfo {author} {\bibfnamefont {Y.}~\bibnamefont {Mei}}, \bibinfo
  {author} {\bibfnamefont {Y.}~\bibnamefont {Zuo}}, \bibinfo {author}
  {\bibfnamefont {Y.}~\bibnamefont {Zhai}}, \bibinfo {author} {\bibfnamefont
  {J.}~\bibnamefont {Li}}, \bibinfo {author} {\bibfnamefont {J.}~\bibnamefont
  {Wen}}, \ and\ \bibinfo {author} {\bibfnamefont {S.}~\bibnamefont {Du}},\
  }\bibfield  {title} {\enquote {\bibinfo {title} {Anti-parity-time symmetric
  optical four-wave mixing in cold atoms},}\ }\href@noop {} {\bibfield
  {journal} {\bibinfo  {journal} {Phys. Rev. Lett.}\ }\textbf {\bibinfo
  {volume} {123}},\ \bibinfo {pages} {193604} (\bibinfo {year}
  {2019})}\BibitemShut {NoStop}%
\bibitem [{\citenamefont {Luo}, \citenamefont {Zhang},\ and\ \citenamefont
  {Du}(2022)}]{APTSQZ}%
  \BibitemOpen
  \bibfield  {author} {\bibinfo {author} {\bibfnamefont {X.-W.}\ \bibnamefont
  {Luo}}, \bibinfo {author} {\bibfnamefont {C.}~\bibnamefont {Zhang}}, \ and\
  \bibinfo {author} {\bibfnamefont {S.}~\bibnamefont {Du}},\ }\bibfield
  {title} {\enquote {\bibinfo {title} {Quantum squeezing and sensing with
  pseudo-anti-parity-time symmetry},}\ }\href@noop {} {\bibfield  {journal}
  {\bibinfo  {journal} {Phys. Rev. Lett.}\ }\textbf {\bibinfo {volume} {128}},\
  \bibinfo {pages} {173602} (\bibinfo {year} {2022})}\BibitemShut {NoStop}%
\bibitem [{\citenamefont {Lawrie}\ \emph {et~al.}(2019)\citenamefont {Lawrie},
  \citenamefont {Lett}, \citenamefont {Marino},\ and\ \citenamefont
  {Pooser}}]{quant_sense:pooser:2018}%
  \BibitemOpen
  \bibfield  {author} {\bibinfo {author} {\bibfnamefont {B.~J.}\ \bibnamefont
  {Lawrie}}, \bibinfo {author} {\bibfnamefont {P.~D.}\ \bibnamefont {Lett}},
  \bibinfo {author} {\bibfnamefont {A.~M.}\ \bibnamefont {Marino}}, \ and\
  \bibinfo {author} {\bibfnamefont {R.~C.}\ \bibnamefont {Pooser}},\ }\bibfield
   {title} {\enquote {\bibinfo {title} {Quantum sensing with squeezed light},}\
  }\href@noop {} {\bibfield  {journal} {\bibinfo  {journal} {ACS Photonics}\
  }\textbf {\bibinfo {volume} {6}},\ \bibinfo {pages} {1307--1318} (\bibinfo
  {year} {2019})}\BibitemShut {NoStop}%
\bibitem [{\citenamefont {Anderson}\ \emph {et~al.}(2017)\citenamefont
  {Anderson}, \citenamefont {Schmittberger}, \citenamefont {Gupta},
  \citenamefont {Jones},\ and\ \citenamefont {Lett}}]{PhysRevA.95.063843Lett}%
  \BibitemOpen
  \bibfield  {author} {\bibinfo {author} {\bibfnamefont {B.~E.}\ \bibnamefont
  {Anderson}}, \bibinfo {author} {\bibfnamefont {B.~L.}\ \bibnamefont
  {Schmittberger}}, \bibinfo {author} {\bibfnamefont {P.}~\bibnamefont
  {Gupta}}, \bibinfo {author} {\bibfnamefont {K.~M.}\ \bibnamefont {Jones}}, \
  and\ \bibinfo {author} {\bibfnamefont {P.~D.}\ \bibnamefont {Lett}},\
  }\bibfield  {title} {\enquote {\bibinfo {title} {Optimal phase measurements
  with bright- and vacuum-seeded su(1,1) interferometers},}\ }\href {\doibase
  10.1103/PhysRevA.95.063843} {\bibfield  {journal} {\bibinfo  {journal} {Phys.
  Rev. A}\ }\textbf {\bibinfo {volume} {95}},\ \bibinfo {pages} {063843}
  (\bibinfo {year} {2017})}\BibitemShut {NoStop}%
\bibitem [{\citenamefont {Prajapati}, \citenamefont {Niu},\ and\ \citenamefont
  {Novikova}(2021)}]{ziqi19}%
  \BibitemOpen
  \bibfield  {author} {\bibinfo {author} {\bibfnamefont {N.}~\bibnamefont
  {Prajapati}}, \bibinfo {author} {\bibfnamefont {Z.}~\bibnamefont {Niu}}, \
  and\ \bibinfo {author} {\bibfnamefont {I.}~\bibnamefont {Novikova}},\
  }\bibfield  {title} {\enquote {\bibinfo {title} {Quantum-enhanced two-photon
  spectroscopy using two-mode squeezed light},}\ }\href@noop {} {\bibfield
  {journal} {\bibinfo  {journal} {Optics Letters}\ }\textbf {\bibinfo {volume}
  {46}},\ \bibinfo {pages} {1800--1803} (\bibinfo {year} {2021})}\BibitemShut
  {NoStop}%
\bibitem [{\citenamefont {Jiang}, \citenamefont {Mei},\ and\ \citenamefont
  {Du}(2023)}]{LangevinPRA2023}%
  \BibitemOpen
  \bibfield  {author} {\bibinfo {author} {\bibfnamefont {Y.}~\bibnamefont
  {Jiang}}, \bibinfo {author} {\bibfnamefont {Y.}~\bibnamefont {Mei}}, \ and\
  \bibinfo {author} {\bibfnamefont {S.}~\bibnamefont {Du}},\ }\bibfield
  {title} {\enquote {\bibinfo {title} {Quantum langevin theory for two coupled
  phase-conjugated electromagnetic waves},}\ }\href {\doibase
  10.1103/PhysRevA.107.053703} {\bibfield  {journal} {\bibinfo  {journal}
  {Phys. Rev. A}\ }\textbf {\bibinfo {volume} {107}},\ \bibinfo {pages}
  {053703} (\bibinfo {year} {2023})}\BibitemShut {NoStop}%
\bibitem [{\citenamefont {McCormick}\ \emph {et~al.}(2008)\citenamefont
  {McCormick}, \citenamefont {Marino}, \citenamefont {Boyer},\ and\
  \citenamefont {Lett}}]{lettPRA08}%
  \BibitemOpen
  \bibfield  {author} {\bibinfo {author} {\bibfnamefont {C.~F.}\ \bibnamefont
  {McCormick}}, \bibinfo {author} {\bibfnamefont {A.~M.}\ \bibnamefont
  {Marino}}, \bibinfo {author} {\bibfnamefont {V.}~\bibnamefont {Boyer}}, \
  and\ \bibinfo {author} {\bibfnamefont {P.~D.}\ \bibnamefont {Lett}},\
  }\bibfield  {title} {\enquote {\bibinfo {title} {Strong low-frequency quantum
  correlations from a four-wave-mixing amplifier},}\ }\href {\doibase
  10.1103/PhysRevA.78.043816} {\bibfield  {journal} {\bibinfo  {journal} {Phys.
  Rev. A}\ }\textbf {\bibinfo {volume} {78}},\ \bibinfo {pages} {043816}
  (\bibinfo {year} {2008})}\BibitemShut {NoStop}%
\bibitem [{\citenamefont {Boyer}\ \emph {et~al.}(2008)\citenamefont {Boyer},
  \citenamefont {Marino}, \citenamefont {Pooser},\ and\ \citenamefont
  {Lett}}]{lettSci08}%
  \BibitemOpen
  \bibfield  {author} {\bibinfo {author} {\bibfnamefont {V.}~\bibnamefont
  {Boyer}}, \bibinfo {author} {\bibfnamefont {A.~M.}\ \bibnamefont {Marino}},
  \bibinfo {author} {\bibfnamefont {R.~C.}\ \bibnamefont {Pooser}}, \ and\
  \bibinfo {author} {\bibfnamefont {P.~D.}\ \bibnamefont {Lett}},\ }\bibfield
  {title} {\enquote {\bibinfo {title} {Entangled images from four-wave
  mixing},}\ }\href {http://link.aps.org/doi/10.1103/PhysRevA.78.043816}
  {\bibfield  {journal} {\bibinfo  {journal} {Science}\ }\textbf {\bibinfo
  {volume} {321}},\ \bibinfo {pages} {544--547} (\bibinfo {year}
  {2008})}\BibitemShut {NoStop}%
\bibitem [{\citenamefont {Wu}\ \emph {et~al.}(2019)\citenamefont {Wu},
  \citenamefont {Schmittberger}, \citenamefont {Brewer}, \citenamefont
  {Speirs}, \citenamefont {Jones},\ and\ \citenamefont {Lett}}]{LettOE2019}%
  \BibitemOpen
  \bibfield  {author} {\bibinfo {author} {\bibfnamefont {M.-C.}\ \bibnamefont
  {Wu}}, \bibinfo {author} {\bibfnamefont {B.~L.}\ \bibnamefont
  {Schmittberger}}, \bibinfo {author} {\bibfnamefont {N.~R.}\ \bibnamefont
  {Brewer}}, \bibinfo {author} {\bibfnamefont {R.~W.}\ \bibnamefont {Speirs}},
  \bibinfo {author} {\bibfnamefont {K.~M.}\ \bibnamefont {Jones}}, \ and\
  \bibinfo {author} {\bibfnamefont {P.~D.}\ \bibnamefont {Lett}},\ }\bibfield
  {title} {\enquote {\bibinfo {title} {Twin-beam intensity-difference squeezing
  below 10 hz},}\ }\href {\doibase 10.1364/OE.27.004769} {\bibfield  {journal}
  {\bibinfo  {journal} {Opt. Express}\ }\textbf {\bibinfo {volume} {27}},\
  \bibinfo {pages} {4769--4780} (\bibinfo {year} {2019})}\BibitemShut {NoStop}%
\bibitem [{\citenamefont {Prajapati}\ and\ \citenamefont
  {Novikova}(2019)}]{Prajapati:19}%
  \BibitemOpen
  \bibfield  {author} {\bibinfo {author} {\bibfnamefont {N.}~\bibnamefont
  {Prajapati}}\ and\ \bibinfo {author} {\bibfnamefont {I.}~\bibnamefont
  {Novikova}},\ }\bibfield  {title} {\enquote {\bibinfo {title}
  {Polarization-based truncated su(1,1) interferometer based on four-wave
  mixing in rb vapor},}\ }\href {\doibase 10.1364/OL.44.005921} {\bibfield
  {journal} {\bibinfo  {journal} {Opt. Lett.}\ }\textbf {\bibinfo {volume}
  {44}},\ \bibinfo {pages} {5921--5924} (\bibinfo {year} {2019})}\BibitemShut
  {NoStop}%
\bibitem [{\citenamefont {Prajapati}\ \emph {et~al.}(2019)\citenamefont
  {Prajapati}, \citenamefont {Super}, \citenamefont {Lanning}, \citenamefont
  {Dowling},\ and\ \citenamefont {Novikova}}]{Prajapati19OL}%
  \BibitemOpen
  \bibfield  {author} {\bibinfo {author} {\bibfnamefont {N.}~\bibnamefont
  {Prajapati}}, \bibinfo {author} {\bibfnamefont {N.}~\bibnamefont {Super}},
  \bibinfo {author} {\bibfnamefont {N.~R.}\ \bibnamefont {Lanning}}, \bibinfo
  {author} {\bibfnamefont {J.~P.}\ \bibnamefont {Dowling}}, \ and\ \bibinfo
  {author} {\bibfnamefont {I.}~\bibnamefont {Novikova}},\ }\bibfield  {title}
  {\enquote {\bibinfo {title} {Optical angular momentum manipulations in a
  four-wave mixing process},}\ }\href {\doibase 10.1364/OL.44.000739}
  {\bibfield  {journal} {\bibinfo  {journal} {Opt. Lett.}\ }\textbf {\bibinfo
  {volume} {44}},\ \bibinfo {pages} {739--742} (\bibinfo {year}
  {2019})}\BibitemShut {NoStop}%
\end{thebibliography}%
\end{document}